\documentclass[a4paper,12pt]{article}
\usepackage[utf8]{inputenc}
\usepackage{cite}
\usepackage[margin=1.3 in]{geometry}
\usepackage{braket}
\usepackage{amsmath}
\usepackage{amsfonts}
\usepackage{hyperref}
\usepackage{graphicx}
\usepackage{comment}
\usepackage{xcolor}
\usepackage{amssymb}
\usepackage{authblk}
\DeclareUnicodeCharacter{202F}{\,}
\DeclareUnicodeCharacter{03BC}{\,} 

\title{\huge Dark Matter and Muon $g-2$ Anomaly via Scale Symmetry Breaking}
\author{\small Parsa Ghorbani}
\date{}
\affil{\it Physics Department, Faculty of Science, Ferdowsi University of Mashhad, Iran}

\newcommand{\be}{\begin{equation}}
\newcommand{\ee}{\end{equation}}
\newcommand{\bs}{\begin{split}}
\newcommand{\es}{\end{split}}

\begin{document}

\maketitle
\begin{abstract}
The Standard Model (SM) without the Higgs mass term is scale invariant. Gildener and Weinberg generalized the scale invariant standard model (SISM) by including the multiplication of scalars in quartic forms. They pointed out that along the flat direction only one scalar -called the scalon- is classically massless and all other scalars are massive. Here we choose a SISM with one scalon and one heavy scalar and extend that further respecting the scale invariance by a vector-like lepton (VLL). By an appropriate choice of the flat direction, the heavy scalar enjoys the $\mathbb{Z}_2$ symmetry and is assumed as DM particle. The scalon connects the visible and dark sector via the Higgs-portal and by interacting with both the muon lepton and the VLL. The VLL is charged under $U(1)_Y$ and interacts with $\gamma/Z$ bosons. We show that the model correctly accounts for the observed dark matter (DM) relic abundance in the universe, while naturally evading the current and future bounds from direct detection (DD) experiments. Moreover, the model is capable to explain the $(g-2)_\mu$ anomaly observed in Fermilab. 
We also show a feature in SISM scenarios which is not present in other Higgs-portal models; despite having the Higgs-portal term $|H|^2 s^2$ ($s$ being the scalon) in SISM, the effective potential after the electroweak symmetry breaking lacks an important expected vertex $h s^2$. This property immediately forbids the tree-level invisible Higgs decay $h\to ss$ and the one-loop Higgs decay $h\to \mu^+ \mu^-$.

\end{abstract}

\newpage

\tableofcontents
\section{Introduction}
Recently the Fermilab National Accelerator Laboratory (FNAL) announced the
results of an improved measurement of the muon $g-2$ magnetic moment \cite{Muong-2:2021ojo}
based on the previous measurement E821 in Brookhaven National Laboratory (BNL) \cite{Muong-2:2006rrc}. The new results indicate a $3.3 \sigma$ discrepancy compared
with the muon $g-2$ prediction in the Standard Model (SM). This anomaly might
be an evidence for new physics (NP). There are different avenues to interpret the
muon $g-2$ anomaly by introducing new particles which interact with the muon
lepton $\mu$ in the SM. There are various extensions of the SM for the explanation
of the muon anomaly. Some examples are: lepton-flavor violating $U(1)'$ extension\cite{Murakami:2001cs}, the aligned
2HDM \cite{Cherchiglia:2016eui}, scalars and vectors that interacts with the leptons and the quarks in
the SM named as leptoquark models \cite{Chakraverty:2001yg,Queiroz:2014zfa,Biggio:2016wyy,Popov:2016fzr} and models with vector-like leptons
(VLL) \cite{Kannike:2011ng,Dermisek:2013gta,Poh:2017tfo, Saez:2021qta, Crivellin:2021rbq,Lee:2021gnw,Barman:2018jhz,Borah:2021khc,Ghorbani:2021yiw}. For reviews on theory and experiment of muon anomalous magnetic moment and possible beyond the standard model (BSM) scenarios see
\cite{Miller:2007kk,Jegerlehner:2009ry,Athron:2021iuf}.
The SISM has also been exploited to explain the observed DM relic density \cite{Steele:2013fka, Ghorbani:2015xvz,YaserAyazi:2018lrv}.

It was proposed in \cite{Bardeen:1995kv} and believed later in the literature e.g. \cite{Meissner:2006zh,Foot:2007iy,Shaposhnikov:2008xi} that the scale invariance might cure the hierarchy problem in the SM. However, this claim was criticized in \cite{MarquesTavares:2013szc} and later the idea was investigated more in \cite{Pelaggi:2017wzr}.      

The scale symmetry is broken by radiative corrections {\it \`{a} la} Coleman and Weinberg \cite{Coleman:1973jx}. Consequently the Higgs gains mass and the electroweak symmetry breaking takes place. To obtain the correct physical properties of the SM, it is necessary to include more than two singlet scalars as prescribed by the Gildener and Weinberg \cite{Gildener:1976ih}. In this article we will
extend the SISM by an extra Dirac vector-like lepton coupled to the SM muon and charged under the SM $U(1)_Y$ while preserving the classical scale invariance. We will investigate whether the aforementioned extended SISM including with two extra real singlet scalars in addition
to the Higgs field, can be used to explain the FNAL muon anomaly $a_\mu=(g-2)_\mu/2$, as well as the observed DM relic density in the universe.

Although the theory is considerably restricted due to the classical
scale symmetry, nevertheless we will show that it is capable of accommodating the dark matter (DM)
relic density and the direct detection (DD) bounds. According to Gildener and Weinberg \cite{Gildener:1976ih}, regardless of the number of extra scalars we include in the SISM, there is always a classically massless scalar along the flat direction called the scalon. All other scalars in the generic model will be massive. In scalar sector of the current model, the DM candidate among two extra scalars, is the heavy one which enjoys a $\mathbb{Z}_2$ symmetry and therefore is stable. The other scalar, i.e. the scalon plays the role of the SM-DM mediator through two portals; the Higgs portal and the muon-VLL portal. 
In any Higgs-portal model there is a term as, 
\begin{equation}
\mathcal{L} \supset  \lambda_\text{hs} |H|^2 s^2
\end{equation}
where $s$ is an extra scalar. Because the Higgs takes non-zero vacuum expectation value (VEV), the vertex $h s^2$ with $h$ being the neutral component of the Higgs doublet, is ubiquities in all Higgs-portal models. 
A remarkable feature of the scale invariant models which has not drawn attention in the literature is that although by radiative correction both the scalon and the Higgs field take non-zero VEV, however along the flat direction, the vertex $hs^2$ is absent. This in turn forbids some tree-level or one-loop signals in experiments, e.g. the tree-level invisible Higgs decay to the scalar $s$, or the one-loop Higgs decay to the muon pair. 

The paper is arranged as the following. In section \ref{model} we will set up the scale invariant model with introducing a VLL coupled to the SM muon.
An appropriated choice of the flat direction is used for which the scalon gains mass by radiative corrections. Then the effective potential after the symmetry breaking is obtained and it is shown that some terms which are ubiquitous in Higgs-portal models, are missing in the SISM. In section \ref{const} we will impose all the relevant constraints; the observed DM relic density in the universe, the bounds from DD experiments, the muon anomaly constraint, and limits on the VLL mass from the soft lepton searches and the LHC 13-TeV. The numerical results are presented in section \ref{res}. we conclude in section \ref{con}.

\section{Scale Invariant Model}\label{model}
The presence of the Higgs mass term in the Higgs potential of the SM is essential for the model to 
undergo an electroweak phase transition from zero vacuum expectation value (VEV) for the Higgs field to the current observed non-zero VEV, which in turn provides mass for other particles of the SM. However, the very same term would
be as well the source of the Higgs hierarchy problem.  The SM without the Higgs mass term is classically scale invariant. The Higgs potential then consists of only a quartic term such as $V=1/4 \lambda_\text{h}h^4$, where $h$ stands for the neutral component of the Higgs doublet. The potential can be extended by more scalars while respecting the scale invariance, for instance the potential with an extra scalar $s$ will be of the form $ V=1/4\lambda_\text{h} h^4+1/2\lambda_\text{hs} h^2 s^2+1/4 \lambda_\text{s} s^4$. The general form of the scale invariant Higgs potential for the SM was suggested by Gildener and Weinberg \cite{Gildener:1976ih}. 
With the current precision measurements in the SM, e.g. for the top quark mass, the gauge boson masses and the Higgs mass, a scale invariant model must possess at least two extra scalars among which one scalar is classically massless dubbed {\it scalon} that becomes massive through radiative corrections. All other scalars including the Higgs field are classically massive. For more details on the SISM see e.g. \cite{Alexander-Nunneley:2010tyr}. Extra degrees of freedom in SISM models, if stable, can be candidates of the DM (see e.g. \cite{Ghorbani:2015xvz}). 

\subsection{Vector-Like Lepton}
In order to investigate the muon anomaly in the SISM framework we add a vector-like lepton to the model. The new fermion interacts with the muon in the SM and with the scalon field
\be \label{lagr}
\mathcal{L} \supset
\kappa \left( s \bar\mu_R \psi_L + h.c. \right)+ y s \bar\psi \psi\\
\ee
where the coupling $\kappa$ and the Yukawa coupling $y$ are real. The Dirac fermion $\psi$ is singlet under the SM 
$SU(2)_L$ gauge group, but it is charged under $U(1)_Y$ with $Y=-1$.  
Let us assume that there are only two real singlet scalars $s$ and $\varphi$ in addition to the Higgs doublet $H$. As will be discussed in the next section the scalar's VEV along the flat direction are $v_\varphi=0, v_s\neq 0$ and $v_H=246$ GeV. Therefore, the scalar $\varphi$ enjoys the $\mathbb{Z}_2$ symmetry and can be taken as dark matter candidate. We will see later on that the scalar $s$ with an appropriate choice of the flat direction is the classically massless scalon field.  The scalon plays the role of the DM-SM mediator having interactions with the Higgs field, the muon in the SM, the VLL fermion field $\psi$, and the DM i,e. the heavy scalar $\varphi$. The extended Higgs potential in the model reads
\be\label{lagr1}
\begin{split}
&V =\frac{1}{4}\lambda_\text{h}
|H|^4 + \frac{1}{2} \lambda_\text{hs} |H|^2 s^2  + \frac{1}{4} \lambda_\text{s} s^4+\frac{1}{2}\lambda_{\text{s}\varphi} s^2 \varphi^2  +\frac{1}{4} \lambda_\varphi \varphi^4 \,.
\end{split}
\ee
The term $\lambda_{\text{h}\varphi} h^2\varphi^2$ could also be included in the potential in Eq. (\ref{lagr1}), however we assume that $\lambda_{\text{h}\varphi}$ is vanishing at a given scale $\Lambda$. This would not affect the results in section \ref{const}; after the scale symmetry breaking this term is generated in the potential as seen in Eq. (\ref{veff}).

\subsection{Positivity Condition}

The potential must be bounded from below or equivalently must satisfy the positivity condition to prevent the instability of the vacuum. For potential in Eq. (\ref{lagr1}) including the Higgs and two extra singlet scalars the positivity condition is given by $\lambda_\text{h}>0 , \lambda_\text{s}>0 , \lambda_\varphi>0$ and
\be 
\begin{split}
&\left(  \lambda_\text{hs}<0 \wedge \lambda_{\text{s}\varphi}>0  \wedge \lambda_\text{h} \lambda_\text{s} \geq \lambda^2_\text{hs} \right) 
\vee \left(  \lambda_\text{hs}<0 \wedge \lambda_{\text{s}\varphi}<0 \wedge \lambda_\text{h} \lambda_\text{s} \geq \lambda^2_\text{hs} + \frac{\lambda_\text{h}}{\lambda_\varphi} \lambda^2_{\text{s}\varphi}\right)\\
& \vee \left( \lambda_\text{hs}>0 \wedge \lambda_{\text{s}\varphi}>0 \right) \vee
\left(  \lambda_\text{hs}>0 \wedge \lambda_{\text{s}\varphi}<0  \wedge \lambda_\text{s} \lambda_\varphi > \lambda^2_{\text{s}\varphi} \right).
 \end{split}
\ee
However, along the flat direction which will be discussed in the next section before Eq. (\ref{flat1}), we must have $\lambda_\text{hs}<0$ and $\lambda_\text{h} \lambda_\text{s} = \lambda^2_\text{hs}$. Also from the positivity of the DM mass $m_\varphi$ in Eq. (\ref{hmas}), the positivity condition is simply,  
\be 
\begin{split}
\lambda_\text{h}>0 \wedge \lambda_\text{s}>0 \wedge \lambda_\varphi>0 \wedge \lambda_\text{hs}<0 \wedge \lambda_{\text{s}\varphi}>0  \wedge \lambda_\text{h} \lambda_\text{s} = \lambda^2_\text{hs}.
 \end{split}
\ee

\subsection{Flat Direction and Scale Symmetry Breaking} \label{sb}

In general the flat direction $\text{\bf n}$ is defined such that along which the tree level potential  and the potential at minimum are vanishing
\be \label{flatdir}
V(h,s)|_\text{\bf n}=V_\text{min}(h,s)|_\text{\bf n}=0\,.
\ee
The flat direction for the choice of the VEVs mentioned above for the scalar fields $h,s,\varphi$ is $(n_\text{h},n_\text{s},n_\varphi)$ with $n_\varphi=0$.
Assuming $ h=  n_\text{h}  \phi  \equiv \sin(\alpha)\phi$ and  $ s= n_\text{s} \phi \equiv \cos(\alpha)\phi$ the 2-dimensional flat direction from Eq. (\ref{flatdir}) is obtained as, 
\be\label{flat1}
\frac{n^2_\text{h}}{n^2_\text{s}}\equiv\frac{\braket{h}^2}{\braket{s}^2}=-\frac{\lambda_\text{s}}{\lambda_\text{hs}}=-\frac{\lambda_\text{hs}}{\lambda_\text{h}}
\ee
which implies $\lambda^2_\text{hs}-\lambda_\text{h} \lambda_\text{s}=0$ and $\lambda_\text{hs}<0$. This 
is equivalent to having, 
\be\label{flat2}
n^2_\text{h}\equiv \sin^2(\alpha)= \frac{\lambda_\text{s}}{\lambda_\text{s}-\lambda_\text{hs}} \hspace{2cm}
n^2_\text{s}\equiv\cos^2(\alpha)= \frac{-\lambda_\text{hs}}{\lambda_\text{s}-\lambda_\text{hs}}
\ee
that satisfies $n^2_\text{h}+n^2_\text{s}=1$.
The VEV $v_\phi$ is related to $v_\text{h}$ and $v_\text{s}$ through $v_\text{h}=n_\text{h} v_\phi$ and $v_\text{s}= n_\text{s} v_\phi$ respectively which implies $v^2_\phi=v^2_\text{h}+v^2_\text{s}$. \\
The Hessian matrix along the flat direction in Eq. (\ref{flat1}) is given by
\be\label{hess1}
\mathcal{H}= -2 \lambda_\text{hs}\begin{pmatrix}
v^2_\text{s} & - v_\text{h} v_\text{s}  \\
- v_\text{h} v_\text{s} & v^2_\text{h}
\end{pmatrix}.
\ee
Rotating the space of the scalars $(h,s)$ as $h\to h'=\cos(\alpha) h+\sin(\alpha) s$ and $s\to s'= -\sin(\alpha) h+\cos(\alpha) s$ with the angle $\alpha$ given in Eq. (\ref{flat2}), diagonalizes the Hessian matrix in Eq. (\ref{hess1}). The mass eigenvalues read
\be\label{hmas}
m_{\text{s}'}^2=0 \hspace{2cm}
m_{\text{h}'}^2=-2\lambda_\text{hs}v^2_\phi \hspace{2cm}
m_\varphi^2= \lambda_{\text{s}\varphi} v^2_s
\ee
where the massless field ($s'$) accounts for the pseudo-Goldstone boson of the scale symmetry. The VLL mass depends on the scalon VEV and the Yukawa coupling in the Lagrangian in Eq. (\ref{lagr1})
\be
m_\psi=y   v_\text{s} \,.
\ee
The one-loop effective potential along the flat direction at scale $\Lambda$ as obtained by Gildener and Weinberg \cite{Gildener:1976ih} is given by
\be
V({\bf n \phi})=A\phi^4+B\phi^4\log \frac{\phi^2}{\mu^2}
\ee
where the dimensionless coefficients A and B in the $\overline{\text{MS}}$ renormalization scheme are  
\be
\begin{split}
A=\frac{1}{64\pi^2 v_\phi^2} &\Bigg[ m^4_{\text{h}'}\left(-\frac{2}{3}+\log\frac{m^2_{\text{h}'}}{v^2_\phi}\right) +
m^4_\varphi\left(-\frac{2}{3}+\log\frac{m^2_\varphi}{v^2_\phi}\right) \\
&+ 6 m^4_W \left(-\frac{5}{6}+\log\frac{m^2_W}{v^2_\phi}\right)
+3m^4_Z\left(-\frac{5}{6}+\log\frac{m^2_Z}{v^2_\phi}\right)\\
&-12m^4_t\left(-1+\log\frac{m^2_t}{v^2_\phi}\right)-4m^4_\psi \left(-1+\log\frac{m^2_\psi}{v^2_\phi}\right) \Bigg]
\end{split}
\ee
and 
\be
B=\frac{1}{64\pi^2 v_\phi^2} \left( m^4_{\text{h}'} +m^4_\varphi+6 m^4_W + 3m^4_Z -12 m^4_t - 4 m^4_\psi \right) \,. 
\ee
In order for the effective potential to satisfy the extremum condition along the flat direction we should require $dV({\bf n}\phi)/d\phi=0$ which in turn leads to
\be
\frac{A}{2B}=-\frac{1}{4}-\frac{1}{2}\log \left(\frac{v^2_\phi}{\mu^2} \right)
\ee
so that the effective potential is simplified as 
\be
V({\bf n \phi})=B\phi^4 \left( \log\frac{\phi^2}{v^2_\phi}-\frac{1}{2} \right)\,.
\ee
The classically massless scalon obtains a mass {\it \`{a} la} Coleman and Weinberg
\be \label{smas}
m^2_{\text{s}'}=\frac{1}{8\pi^2 v^2_\phi}\left( m^4_{\text{h}'} +m^4_\varphi+6 m^4_W + 3m^4_Z -12 m^4_t - 4 m^4_\psi   \right)\,. 
\ee
Only the mass of the VLL $m_\psi$, and the DM mass $m_\varphi$ are unknown, however the condition $m^2_{\text{s}'}>0$ relates these two parameters as
\be\label{dmlim}
m^4_\varphi > \left( 316.12 ~\text{GeV} \right)^4+ 4m^4_\psi
\ee
where we have used $m_{\text{h}'}=125.10$ GeV, $m_t=172.76$ GeV, $m_W=80.38$ GeV and $m_Z=91.19$ GeV. As an example, for VLL mass $m_\psi\sim 100$ GeV and DM mass $m_\varphi \sim 500$ GeV, if we assume $v_\text{s}\sim 100$ GeV, we get a scalon mass $m_{\text{s}'}\sim 100$ GeV.

\begin{figure}
\begin{center}
\includegraphics[scale=1]{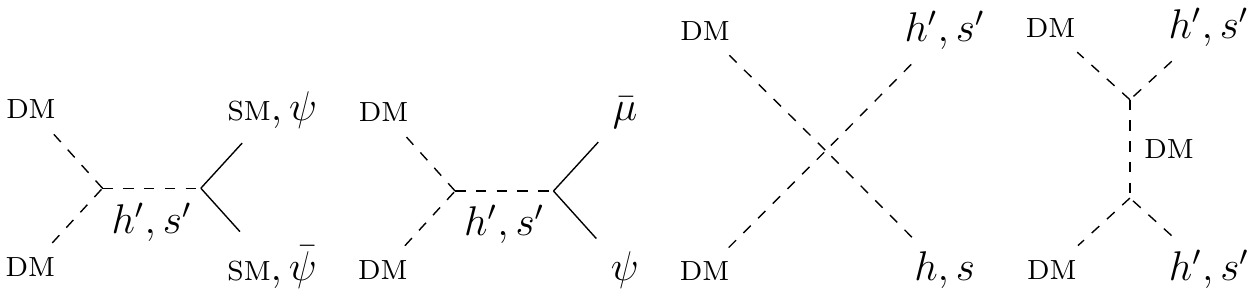}
\caption{Shown are tree-level Feynman diagrams for dark matter annihilation processes.} 
\label{dmann}
\end{center}
\end{figure} 

\subsection{Effective Potential}
After the dimensional transmutation through the radiative corrections and the breakdown of the scale symmetry, the scalon obtains non-zero VEV and mass. The effective potential then contains a mass term for the scalon $s'$ from Eq. (\ref{lagr1}). The scalon having obtained a non-zero VEV is  mixed with the  Higgs field where the mixing angle is given by Eq. (\ref{flat2}). The parameters $(\lambda_\text{h}, \lambda_\text{s}, v_s)$ in the Lagrangian in Eq. (\ref{lagr1}) can be replaced by the parameters $(\alpha,\lambda_\text{hs})$ along the flat direction which in turn implies $\lambda_\text{h}=-v^2_s \lambda_\text{hs}/v^2_h, \lambda_\text{s}=-v^2_h \lambda_\text{hs}/v^2_s$ and $v^2_s=v^2_h \cot(\alpha)$. The free dimensionless parameters in the scalar sector then becomes the set $\{\alpha, \lambda_\text{hs},\lambda_{s\varphi},\lambda_\varphi\}$. The Higgs mass $m_\text{h}$, the scalar DM mass $m_\varphi$, the VLL mass $m_\psi$ and the scalon mass $m_{\text{s}'}$ are known from Eqs. (\ref{hmas}) and (\ref{smas}). 
The effective potential in terms of the new set of free parameters read
\be \label{veff}
\begin{split}
V_\text{}(h',s',\varphi)=&-\lambda_\text{hs} \Big(  \frac{v^2_h}{\sin^2(\alpha)} h'^2 + \frac{2v_h}{\sin(\alpha)\tan(2\alpha)}h'^3+\frac{1}{\tan^2(2\alpha)} h'^4 \\
&+h'^2 s'^2 
+\frac{2v_h}{\sin(\alpha)}h'^2 s' +\frac{2}{\tan(2\alpha)}h'^3 s' \Big) \\
&+\lambda_{s\varphi}\Big( \frac{v_h^2 }{2 \tan^2(\alpha)} \varphi^2
+\frac{1}{2} \cos^2(\alpha) s'^2 \varphi^2 
+\frac{1}{2} \sin^2(\alpha) h'^2 \varphi^2\\
&+\frac{v_h \cos(\alpha)}{\tan(\alpha)}   s'\varphi^2 -v_h \cos(\alpha) h' \varphi^2 -\frac{1}{2} \sin(2\alpha)  h' s' \varphi^2 \Big)+\frac{1}{4}\lambda_\varphi \varphi^4
\end{split}
\ee
The remarkable feature of the effective potential in Eq. (\ref{veff}) is that it lacks some terms that are ubiquitous in all Higgs-portal models with $v_\text{h}\neq 0$ and $v_\text{s}\neq 0$: the vertices $h' s'^2$  and $h' s'^3$ are missing. The absence of these vertices forbids some phenomenological signals in experiments. The relevant examples will be discussed in subsection \ref{hdec}. 

Using $\lambda^2_\text{hs}=\lambda_\text{h} \lambda_\text{s}$, the $(h,s)$ part of the potential takes the form $V(h,s)= \left(\lambda_\text{h} h^2 + \lambda_\text{hs} s^2 \right)^2/ 4\lambda_\text{h}$. In terms of the rotated fields, the potential becomes $V(h',s')= h'^2 \left[ h' (1-\tan^2\alpha) +s'  \tan\alpha   \right]^2/4\lambda_\text{h} $ which is vanishing for $h'=0$ as it should be along the flat direction. This evidently explains the absence of the term $h' s'^2$.

\begin{figure}
    \centering
    \includegraphics[scale=.65]{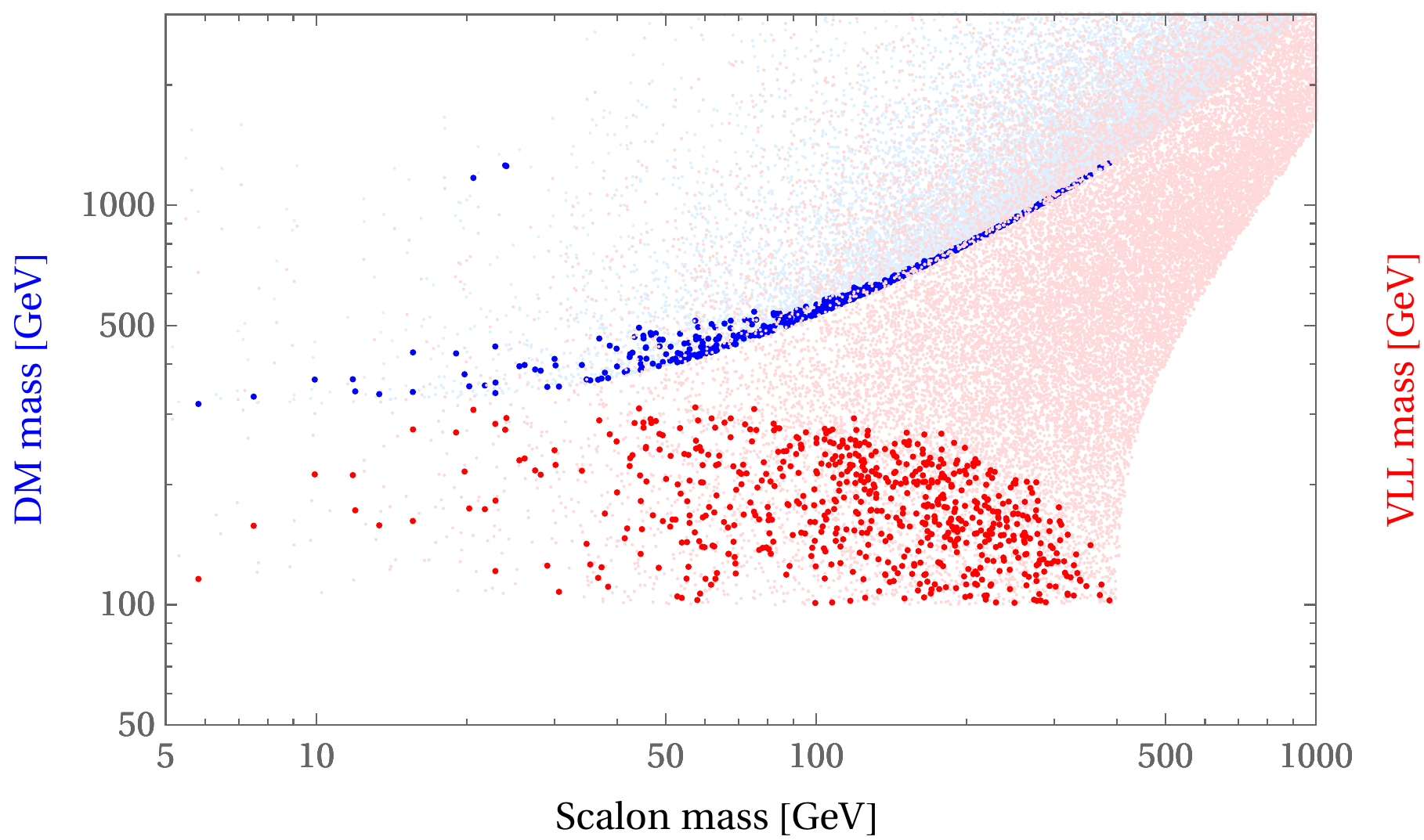}
    \caption{The scalon mass is shown against the DM mass (blue) and the VLL mass (red) where the correct DM relic density is taken into account. The regions with highlighted colors respect also the constraint from the muon $a_\mu$ anomaly.}
    \label{s-dm-vll}
\end{figure}

\section{Constraints}\label{const}
In this section we examine the model by several constraints such as DM relic density, direct detection bounds, muon anomalous magnetic moment and different Higgs decay channels.

\subsection{Dark Matter Relic Density}\label{dmrelic}
The scalar $\varphi$ is stable due to the $\mathbb{Z}_2$ symmetry and is taken as the DM particle in our scale invariant model. 
The scalon $s'$ plays the role of the mediator between the DM and the SM sectors. All possible DM annihilation channels are shown in Fig. \ref{dmann}. As seen in Fig. \ref{dmann}, the DM particle annihilates into SM fermions including heavy quarks top $t$ and bottom $b$, the muon lepton $\mu$, the VLL $\psi$, and the Higgs $h'$ (or the scalon $s'$) through the $s$-channel. The DM annihilates as well into the Higgs and the scalon via the $t$-channel. Note that the DM annihilation may occur in a $2\to 3$ process if the bremsstrahlung radiation is produced from a fermion in the final state. As such contributions are negligible in the DM relic density, we abstained from including the relevant diagrams in Fig. \ref{dmann}. 

The Boltzmann equation which governs the evolution of the DM particle $\varphi$ in the early universe is given by  \cite{Bhattacharya:2013hva}
\be 
\frac{dY}{dt}=- s \braket{\sigma v} \left(Y^2-Y^2_\text{eq} \right)
\ee 
where $Y=n/s $ with $n$ and $s $ being the DM number density and the entropy density of the universe, respectively. 
We make use of the {\tt MicrOMEGAs 5.2.13 } package \cite{Belanger:2010pz} to calculate the DM relic density. The results and the viable space of the parameters are presented in section \ref{res}.

\begin{figure}
\begin{center}
\includegraphics[scale=1]{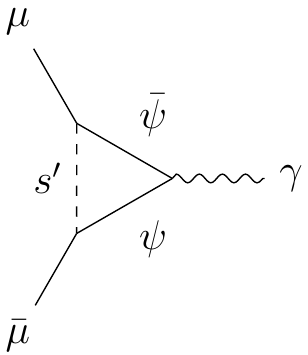}
\caption{The one-loop contribution to muon anomalous magnetic moment through the VLL $\psi$, and the scalon $s'$.}
\end{center}
\label{1loopamu}
\end{figure}

\subsection{Direct Detection}\label{dd}
\begin{figure}
    \centering
    \includegraphics[scale=.92]{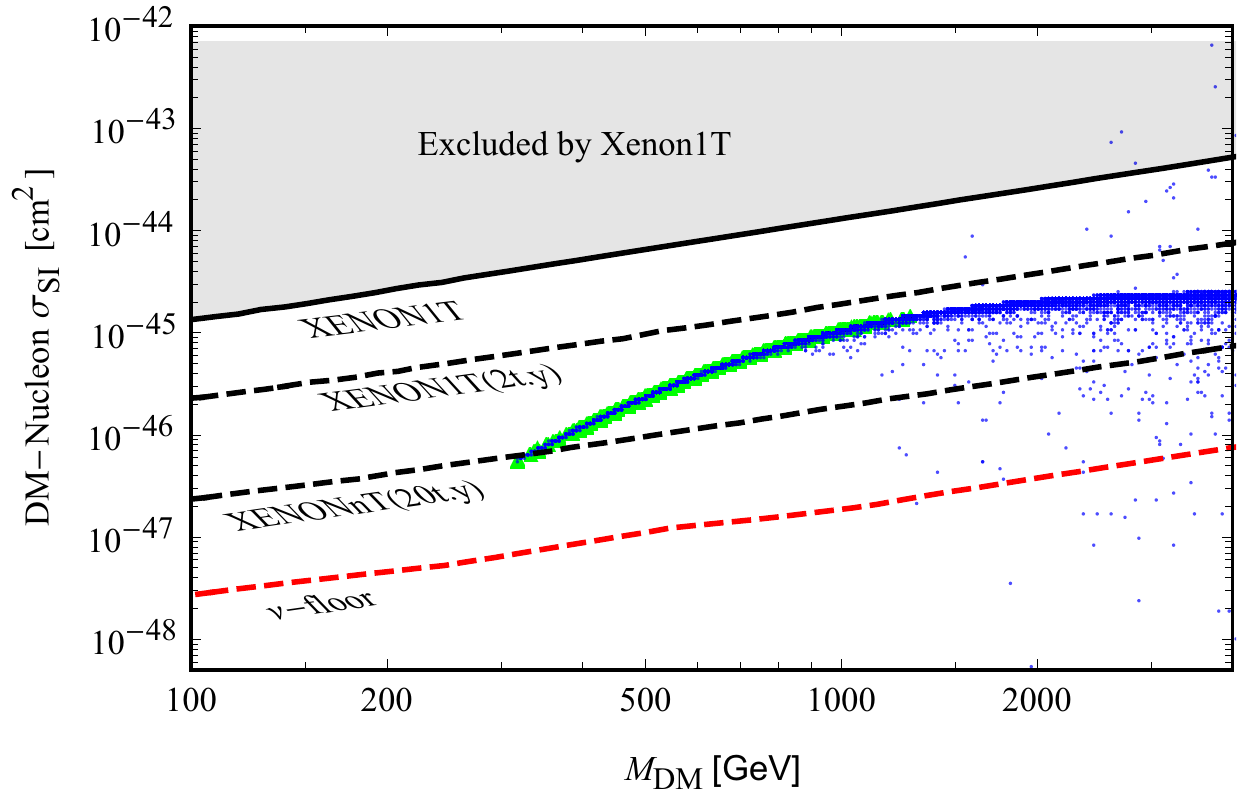}
    \caption{The DM mass vs. spin-independent DM-nucleon cross section is shown. The blue region is the parameter space with the observed DM relic density. The green region in addition explains the muon anomalous magnetic moment. Almost all viable regions for the DM relic density and the muon anomaly lies below the current and future DD bounds.}
    \label{dm-dd}
\end{figure}
The spin-independent SM-nucleon scattering cross section  in the scale invariant model takes contributions from two tree-level Feynman diagrams for $\varphi \varphi \to \bar q q$ with the mediator being either the Higgs particle or the scalon. The effective Lagrangian for DM-nucleon interaction can be written as 
\begin{equation}
\mathcal{L}\supset \mathcal{M} \varphi\varphi \bar q q
\end{equation}
with the effective coupling given by
\begin{equation}
\mathcal{M}=m_\text{q} \lambda_{\text{s}\varphi} \left(\frac{1}{m^2_{\text{h}'}}+\frac{1}{m^2_{\text{s}'}}  \right) \cos^2\alpha.
\end{equation}
Therefore the DM-nucleon cross section becomes
\begin{equation}\label{sigmaSI}
 \sigma_\text{SI}=\frac{\mu^2_N \mathcal{M}^2}{4\pi m^2_\text{DM}} 
\end{equation}
where $\mu_N=m_\text{DM} m_N/(m_\text{DM} + m_N)$ is the DM-nucleon reduced mass. The momentum transfer is taken zero in Eq. (\ref{sigmaSI}) and $m_q \sim f_N m_N$  where $f_N \sim 0.3$ \cite{Cline:2013gha}.
The comparison of the DM-nucleon cross section with the bounds from DD experiments will be made in section \ref{res}.
\subsection{LHC/LEP and Astrophysical Constraints}\label{lhclep}
For the $2\to 3$ coannihilation $\varphi \varphi\to s'+\text{VLL}+ \gamma/Z$  which includes the bremsstrahlung radiation in the final state, the Fermi-LAT constrains the VLL and the scalon masses with $m^2_\psi/m^2_{\text{s}'}<1.2$ \cite{Toma:2013bka}. 

The searches for slepton, neutralino and chargino in the final leptonic states in the 13-TeV collider searches \cite{ATLAS:2018ojr,CMS:2018eqb,CMS-PAS-EXO-16-036,ATLAS:2019lff} excludes part of the $m_\psi$--$m_\text{s}$ plane illustrated by the region shaded light blue in Fig. \ref{lhc-slepton}.

The LEP limit on charged fermion masses \cite{ALEPH:2002gap,DELPHI:2003uqw}, put a lower bound on the mass of the VLL: $m_\psi>100$ GeV. Also the CMS soft lepton searches \cite{CMS:2018kag}, excludes the region shaded cyan in the $m_\psi$--$m_\text{s}$ plane in Fig. \ref{lhc-slepton}. 

\subsection{Muon Anomalous Magnetic Moment}
\label{muonanom}

In the Standard Model the muon magnetic moment stemming from QED, weak and hadronic contributions, is expected to be $a^\text{SM}_\mu=116591810(43)\times 10^{-11}$ \cite{Aoyama:2020ynm}. However, from the E821 experiment at Brookhaven National Laboratory (BNL), the muon magnetic moment turned out to be larger than the SM expected value being $a_\mu^\text{BNL}=11 659 208.0(5.4)(3.3)\times 10^{-10}$ and $\sigma\sim 2$ standard deviation \cite{Muong-2:2006rrc}. Recently an improved measurement of the muon magnetic moment at the Fermi National Laboratory (FNAL) resulted in $a_\mu^\text{FLAN}=116592040(54)×10^{-11}$ which is $\sigma\sim 3.5$ standard deviations larger than the SM prediction \cite{Muong-2:2021ojo}.
Therefore, the experimental measurement for the muon magnetic moment has the following deviation from the SM prediction based on the experimental world average and the White Paper result for the SM in \cite{Aoyama:2020ynm}
\be \label{afnal}
\Delta a_\mu^\text{FNAL}=\left( 25.1\pm 5.9 \right)\times 10^{-10}\,.
\ee
The fermionic part of the Lagrangian in Eq. (\ref{lagr1}) after the scale symmetry breaking and the Higgs-scalon mixing can be expressed as 
\be
\mathcal{L}_\text{VLL} = \kappa \left( \frac{1}{2}s' \bar\mu \psi -\frac{1}{2} s' \bar\mu \gamma^5 \psi + h.c \right) + y s' \bar\psi \psi 
+ y v_s  \bar\psi \psi \,.
\ee
Now the muon magnetic moment correction due to the VLL $\psi$, and the scalon $s'$, in the one-loop Feynman diagram as depicted in the Fig. \ref{1loopamu}
is given by \cite{Hiller:2019mou}
\be\label{a1loop}
\Delta a_\mu=\frac{\kappa^2}{96\pi^2}\frac{m^2_\mu}{m_\psi^2}F\left( \frac{m^2_s}{m^2_\psi} \right) 
\ee
where $F(x)$ is a positive function defined as $F(x)=(2x^3+3x^2-6x^2\log x-6x+1)/(x-1)^4$ with $F(0)=1$. We impose the condition in Eq. (\ref{afnal}) on the one-loop correction to the muon magnetic moment in Eq. (\ref{a1loop}) and will show in section \ref{res} that the model easily accommodates this constraint. 

\begin{figure}
\begin{center}
\includegraphics[scale=1]{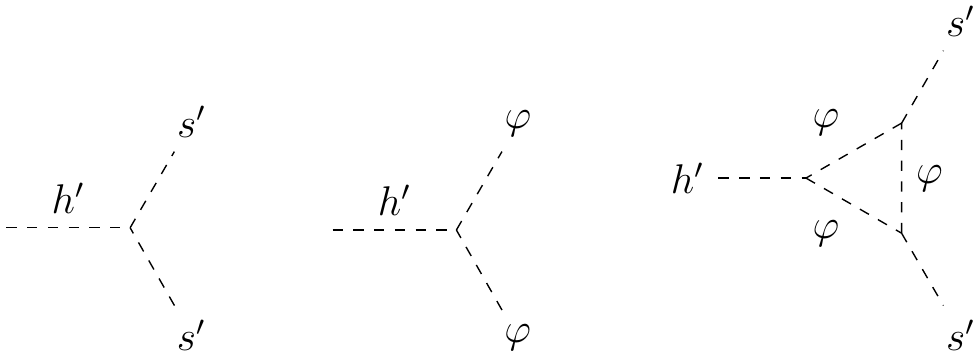}
\caption{The tree-level invisible Higgs decay $h'\to s's'$ is dynamically forbidden. The tree-level $h'\to \varphi \varphi$ and the one-loop $h'\to s's'$ invisible Higgs decays are not allowed kinematically.}\label{hdecay}
\end{center}
\end{figure}

\subsection{Invisible Higgs Decay and Higgs Decay to Muon Pair}\label{hdec}
Although the scalon field $s'$ mixes with the Higgs field, however from Eq. (\ref{veff}) it is evident that the vertex $h' s'^2$ is absent in the effective potential which is not expected in Higgs-portal models as the Higgs field takes non-zero VEV. Therefore, the tree-level invisible Higgs decay channel into the scalon is forbidden in scale invariant models. But the vertex $h' \varphi^2$ does exist in the effective potential in Eq. (\ref{veff}). As seen in Fig. \ref{hdecay}, it implies a tree-level Higgs decay into the heavy scalar field $\varphi$ (DM), however from Eq. (\ref{dmlim}) $m_\text{DM}>m_{\text{h}'}$, so the invisible Higgs decay into the DM is not kinematically allowed. The next leading contribution in the Higgs decay width comes from a one-loop Feynman diagram with the DM particle $\varphi$ in the loop as depicted in Fig. \ref{hdecay}.
However, from the viable space of the parameters after requiring the constraints from the observed DM relic abundance and the muon anomalous magnetic moment shown in section \ref{res}, $m_{\text{s}'}>70>m_{\text{h}'}/2$ GeV. Therefore the one-loop contribution for the invisible Higgs decay into the scalon is also not kinematically allowed. This means that there is no constraint from the invisible Higgs decay width on our model. 

\begin{figure}
\begin{center}
\includegraphics[scale=1]{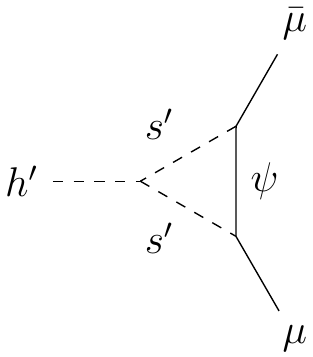}
\caption{The one-loop contribution to the Higgs decay into muon pair, present in Higgs-portal models with VLL-muon interaction. This decay channel is forbidden in our model due to the scale symmetry.} \label{1loophdec}
\end{center}
\end{figure}

A constraint which is present in Higgs-portal models enriched with a VLL coupled to the muon lepton, is the Higgs decay into the muon pairs as shown in Fig. \ref{1loophdec}. Another remarkable feature of the scale invariant model is that such one-loop Feynman diagram is again forbidden due to the absence of the vertex  $h' s'^2$ in the effective potential in Eq. (\ref{veff}), hence no constraint on the model from $h'\to \mu^+ \mu^-$ decay. 

\section{Results}\label{res}
We scanned through the set of free parameters $\lbrace \alpha, m_{\text{h}'}, m_{\text{s}'}, m_\psi, \kappa \rbrace$ by implementing the model in the {\tt MicrOMEGAs 5.2.13} package. The observed DM relic density as discussed in subsection \ref{dmrelic} is to lie in the interval $0.1172<\Omega_\text{DM} h^2<0.1226$ ($h \sim 0.7$ is the normalized Hubble parameter). We also imposed the VLL mass limit $m_\psi>100$ GeV and the condition $m^2_\psi/m^2_{\text{s}'}<1.2$ discussed in subsection \ref{lhclep}. To these the muon $a_\mu$ anomaly condition given in th subsection \ref{muonanom} was added. 

From the calculations it turns out that the scalar field $\varphi$ in the model can be responsible for the whole observed DM relic abundance $\Omega_\text{DM} h^2\sim 0.12$. In Fig. \ref{s-dm-vll}, the mass of the scalon is depicted with respect to the DM mass and the VLL mass; the blue color shows the scalon mass versus the DM mass, while the red color shows the scalon mass versus the VLL mass. The highlighted blue and red regions are the viable space respecting the muon anomaly. 
As seen from Fig. \ref{s-dm-vll}, the scalon can be as light as $5$ GeV for which the DM mass takes values around its lowest masses i.e. $\sim 318$ GeV (and VLL mass being around $100$ GeV) while respecting all the constrains except the DD bound. 
 
In Fig. \ref{dm-dd}, the DM mass versus the spin-independent DM-nucleon cross section is illustrated. It is shown that almost all the viable region depicted in Fig. \ref{s-dm-vll} naturally lies below the current and future bounds from the DD experiments XENON1T and XENONnT. In comparison with the DM-nucleon cross section in the singlet scalar model (see e.g. \cite{Cline:2013gha}), beside the mixing contribution $(1/m^2_{\text{h}'}+1/m^2_{\text{s}'})$, the main difference comes from the coefficient $\cos^4\alpha$ in Eq. (\ref{sigmaSI}). From the viable parameter space of the correct DM relic density and the expected muon anomaly we have $\cos\alpha \sim 0.1$. Therefore, $\sigma_\text{SI}$  in our model is $\mathcal{O}(10^{-2})$ smaller than that of the singlet scalar model.

The region highlighted in green color is the region for which the observed muon anomaly is explained correctly. That is the DM mass range $\sim 318$--$1275$ GeV. The allowed values for the coupling $\kappa$ to account for the correct muon $g-2$ anomaly sits in the range $\sim 1.5$--$4.3$. From Eq. (\ref{a1loop}) it is inferred that the larger values of the coupling $\kappa$ is needed when the larger VLL mass is invoked. The lightest allowed VLL in our scenario is $\sim 100$ GeV which depending on the scalon mass leads approximately to $\kappa\sim 1.5$ in order to accommodate the muon anomaly in Eq. (\ref{afnal}). 
 
\begin{figure}
    \centering
    \includegraphics[scale=.65]{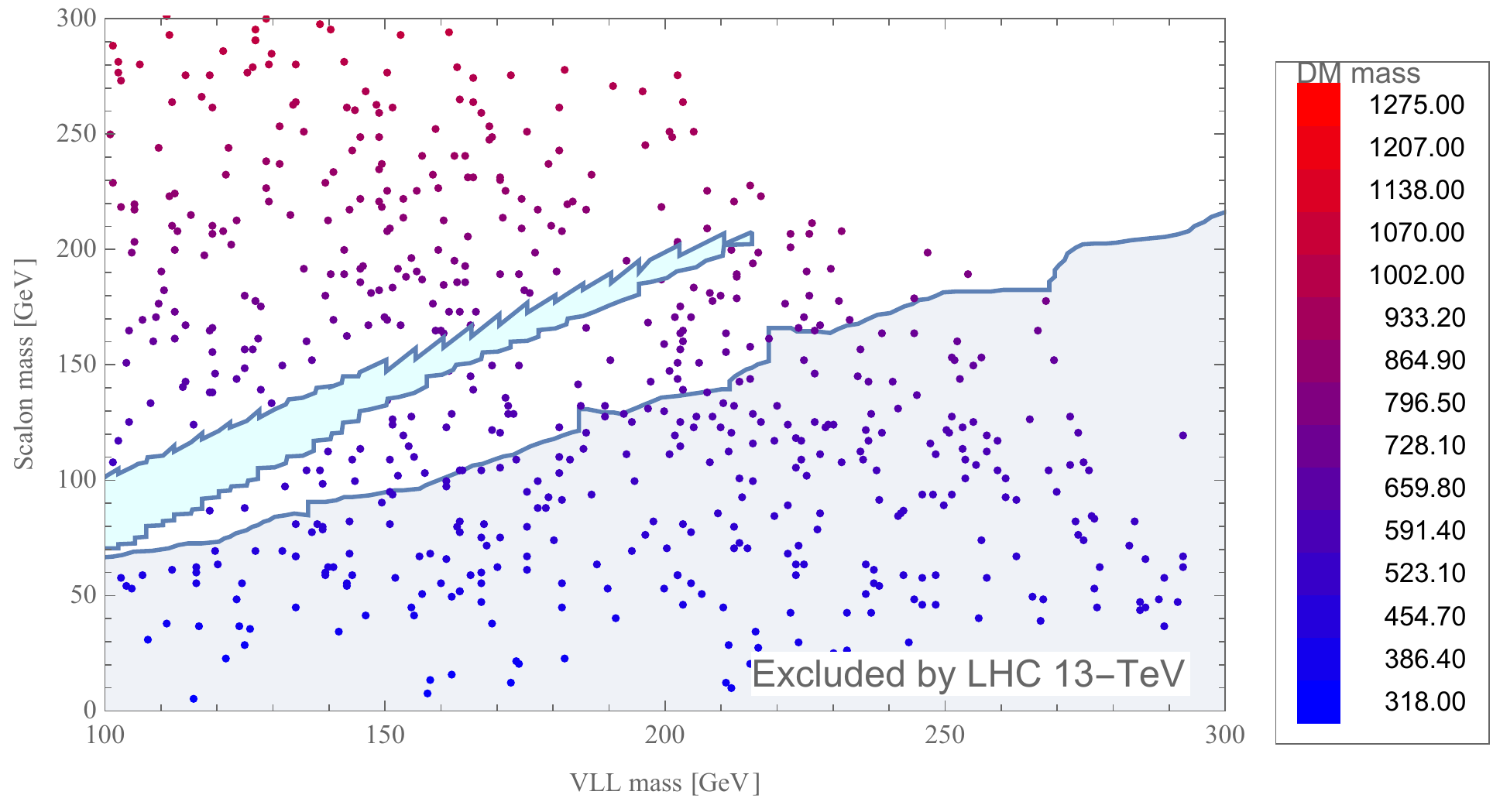}
    \caption{The plot shows the VLL mass against the scalon mass with the DM mass in the color axis. The light blue region is excluded by the 13-TeV LHC and the region shaded cyan is excluded by the soft lepton searches.}
    \label{lhc-slepton}
\end{figure}

Taking into account the LHC/LEP constraints we would impose all the constraints discussed in section \ref{const}. In Fig. \ref{lhc-slepton}, the VLL mass is shown against the scalon mass, with the DM mass shown in color axis, where the LEP limit on the VLL mass $m_\psi>100$ GeV has been considered. The light blue region is excluded from the LHC 13-TeV searches and the region shaded cyan is excluded by the CMS soft lepton searches as discussed in subsection \ref{lhclep}. The regions left in white are allowed after taking into account all the constraints discussed in section \ref{const}.  
From the Fig. \ref{lhc-slepton} the viable region for the scalon mass is $m_{\text{s}'}>70$ GeV which is equivalent to DM viable mass $m_\text{DM} \sim 500$--$1275$ GeV. 

\section{Conclusion}\label{con}
Recently Fermi National Laboratory (FNAL) announced the muon $(g-2)$ anomaly which might be interpreted as a signature of new physics. In this article,  we have considered an extension of the scale invariant standard model (SISM) possessing two real singlet scalar in addition to the Higgs particle and a vector-like lepton (VLL) coupled to the SM muon lepton. We have tried to answer the question whether in such model it is possible to accommodate the DM problem with its relevant astrophysical and experimental constraints as well as the recent muon anomalous magnetic moment. The introduction of the VLL restrict the model from the LHC/LEP and the Fermi-LAT constraints. It should be noted that already the SISM is considerably restricted due to the scale symmetry. Among the two scalars in the model, one is the scalon which gains its mass through radiative corrections {\it \`{a} la} Coleman-Weinberg. The other scalar is heavy and stable due to the $\mathbb{Z}_2$ symmetry and we assume that as the DM particle. We have shown that the model can easily fulfill all the constraints mentioned above including the recent muon $g-2$ anomaly.

We also have showed a feature of SISM models which is absent in other Higgs-portal scenarios in which the term $h^2 s^2$ is always present; although the Higgs field takes non-zero VEV, we have shown that in SISM the vertices $h s^2$ and $h s^3$ are absent in the effective potential. This property in turn makes the tree-level Higgs decay into the scalon and the Higgs decay into the muon pair forbidden.

\section*{Acknowledgement}
I am thankful to Karim Ghorbani and Alessandro Strumia for useful discussions. 

\bibliography{ref.bib}
\bibliographystyle{utphys}
\end{document}